\documentclass[fleqn,usenatbib]{mnras}

\usepackage[T1]{fontenc}

\DeclareRobustCommand{\VAN}[3]{#2}
\let\VANthebibliography\thebibliography
\def\thebibliography{\DeclareRobustCommand{\VAN}[3]{##3}\VANthebibliography}

\usepackage{graphicx}	
\usepackage{amsmath}	
\usepackage{color,soul}
\usepackage{lineno}

\usepackage{newtxtext,newtxmath}

\title[Super-luminous stars in $\gamma$-rays]{A study of super-luminous stars with the \textit{Fermi} Large Area Telescope}

\author[R. de Menezes et al.]{
Raniere de Menezes,$^{1}$\thanks{E-mail: raniere.m.menezes@gmail.com}
Elena Orlando,$^{2,3}$
Mattia Di Mauro,$^{4}$
and Andrew Strong$^{5}$
\\
$^{1}$Universidade de S\~ao Paulo, Departamento de Astronomia, Rua do Mat\~ao, 1226, S\~ao Paulo, SP 05508-090, Brazil\\
$^{2}$INFN Sezione di Trieste and Universit\`a degli Studi di Trieste, Via Valerio 2, 34127 Trieste, Italy \\
$^{3}$Kavli Institute for Particle Astrophysics and Cosmology and Hansen Experimental Physics Laboratory, Stanford University, CA, USA \\
$^{4}$INFN Torino, Physics Department, Via Pietro Giuria 1, 10125 Torino, Italy \\
$^{5}$Max-Planck-Institut f\"ur extraterrestrische Physik, Germany}

\date{Accepted XXX. Received YYY; in original form ZZZ}

\pubyear{2021}

\begin{document}
\label{firstpage}
\pagerange{\pageref{firstpage}--\pageref{lastpage}}
\maketitle

\begin{abstract}
The $\gamma$-ray emission from stars is induced by the interaction of cosmic rays with stellar atmospheres and photon fields. This emission is expected to come in two components: a stellar disk emission, where $\gamma$-rays are mainly produced in atmospheric showers generated by hadronic cosmic rays, and an extended halo emission, where the high density of soft photons in the surroundings of stars create a suitable environment for $\gamma$-ray production via inverse Compton (IC) scattering by cosmic-ray electrons. Besides the Sun, no other disk or halo from single stars has ever been detected in $\gamma$-rays. However, by assuming a cosmic-ray spectrum similar to that observed on Earth, the predicted $\gamma$-ray emission of super-luminous stars, like e.g. Betelgeuse and Rigel, could be high enough to be detected by the \textit{Fermi} Large Area Telescope (LAT) after its first decade of operations. In this work, we use 12 years of \textit{Fermi}-LAT observations along with IC models to study 9 super-luminous nearby stars, both individually and via stacking analysis. Our results show no significant $\gamma$-ray emission, but allow us to restrict the stellar $\gamma$-ray fluxes to be on average $<3.3 \times 10^{-11}$ ph cm$^{-2}$ s$^{-1}$ at a 3$\sigma$ confidence level, which translates to an average local density of electrons in the surroundings of our targets to be less than twice of that observed for the Solar System.
\end{abstract}



\section{Introduction}
\label{sec:intro}

Stars are relatively shy non-thermal sources of $\gamma$-rays with quiescent emission expected to come in two distinct components, both determined by the cosmic-ray spectrum close to the stars. The first component is the stellar disk emission, for which $\gamma$-rays are produced in atmospheric cascades induced by hadronic cosmic rays. There is no theoretical model describing this hadronic emission for stellar disks in general and, even for the Sun, the $\gamma$-ray emission models for this component \citep{seckel1991signatures,li2020simulating,gutierrez2020sun,mazziotta2020_FLUKA_code} cannot explain all observed features \citep{abdo2011fermi,tang2018unexpected}. The second component is the stellar extended photon halo, where leptonic cosmic rays boost soft photons to the $\gamma$-ray domain via inverse Compton (IC) scattering, in good agreement with theoretical predictions \citep{moskalenko2006inverse,orlando2007gamma,orlando2008gamma}.

So far, the only star for which both $\gamma$-ray components have been detected is the Sun. They were first identified by \cite{orlando2008gamma} based on observations with the Energetic Gamma Ray Experiment Telescope \citep{kanbach1989project}, and have been further studied with {\it Fermi} Large Area Telescope (LAT) data \citep{abdo2011fermi,tang2018unexpected,linden2018evidence}. The solar disk flux $> 100$ MeV measured at periods of low solar activity is $\sim 4.6 \times 10^{-7}$ ph cm$^{-2}$ s$^{-1}$, while the extended halo flux integrated over an angular radius of $20^{\circ}$ (hereafter ``elongation angle'') is $\sim 6.8 \times 10^{-7}$ ph cm$^{-2}$ s$^{-1}$. As the IC emission depends on the local number density of soft photons, this component is brighter in the region very close to the star, but even at large elongation angles ($\sim 20^{\circ}$ for the Sun) it can be as intense as the isotropic $\gamma$-ray background \citep{abdo2011fermi}.

Although no isolated stars are listed in the \textit{Fermi}-LAT fourth source catalog \citep[4FGL;][]{abdollahi2020_4FGL,ballet2020_4FGL-DR2}, the predicted IC $\gamma$-ray emission from extended halos of nearby super-luminous stars could in principle be high enough to be detected in $\gamma$-rays. \cite{orlando2007gamma} showed that if we consider a distribution of cosmic-ray electrons similar to that observed in the Solar System, the $\gamma$-ray emission from the halo component of a few nearby stars could exceed $\sim 10^{-9}$ ph cm$^{-2}$ s$^{-1}$, which is well within \textit{Fermi}-LAT sensitivity threshold. The detection of such extended $\gamma$-ray halos can be used as a powerful tool to probe the distribution of cosmic-ray electrons in various locations in the Galaxy. 

Other attempts at exploring the interactions of cosmic rays in the vicinity of stars have been done recently by \cite{riley2019possible}, where the authors explored the possibility of $\gamma$-ray emission from debris disks around main sequence stars, looking for cosmic-ray interactions similar to those occurring on the surface of the Moon \citep{abdo2012_Moon_fermi-LAT}, but no strong evidence for $\gamma$-ray emission was found. In \cite{song2020stacking} the authors looked for $\gamma$-ray emission from stellar flares in 97 nearby ultracool dwarfs and, even after stacking photon counts for the whole sample, no significant emission was found, with a possible exception for the radio star TVLM 513-46546, where there is evidence for pulsed $\gamma$-ray emission.

In this work we present the first systematic study of a population of super-luminous stars in $\gamma$-rays, where we test if an IC $\gamma$-ray emission model for stellar halos is supported by \textit{Fermi}-LAT observations. We use deep $\gamma$-ray observations achieved by stacking \textit{Fermi}-LAT data to constrain the number density of cosmic-ray electrons in the surroundings of our targets. Alternative channels for stellar $\gamma$-ray production, like $\gamma$-rays from annihilation of dark matter particles trapped in the star's gravitational well \citep{niblaeus2019neutrinos,mazziotta2020search}, are not taken into account. This paper is organized as follows. We describe the criteria for selecting our sample in \S \ref{sec:sample} and detail the adopted IC model in \S \ref{sec:IC_models}. The analysis methods are shown in \S \ref{sec:observations}, while in \S \ref{sec:resultados} we present the results, followed by a discussion in \S \ref{sec:discussion} and a summary of our findings in \S \ref{sec:conclusions}.


\section{Sample description}
\label{sec:sample}

Our sample consists of 9 isolated super-luminous nearby stars with the highest expected IC fluxes, as calculated and listed in \cite{orlando2007gamma}. In the following, we summarize the selection criteria as reported in that work. In a first approximation, for a given cosmic-ray electron density, the IC luminosity $L_{IC}$ within a volume surrounding a star is proportional to the radius $r$ with origin at the center of the star multiplied with the optical luminosity of the star, i.e. $L_{IC} \propto r~L_{star}$. The IC flux $F_{IC}$ depends on the star's distance $d$ as the inverse square law, i.e. $F_{IC} \propto L_{IC} / d^2$. As a consequence, for an elongation angle $\alpha \approx r/d$, the IC flux contained within an angle $\alpha$ surrounding the star is roughly $F_{IC} \propto L_{star}~\alpha/d$. Hence the most luminous and nearby stars are the best candidates for giving the largest IC fluxes. More precise formulations as used in our estimates are reported in \cite{orlando2007gamma, orlando2008gamma, orlando2020stellarics}. 

The best candidates for $\gamma$-ray detection with {\it Fermi}-LAT as selected from \cite{orlando2007gamma} have been chosen from an original list containing the 70 most luminous stars in the Hipparcos and Tycho Catalogues \citep{esa1997_Hipp_Tycho_cats} lying up to a distance of 600~pc from the Sun. Based on the sensitivity of \textit{Fermi}-LAT, the following 9 stars have been identified as good candidates for detection in $\gamma$-rays, with expected fluxes $\gtrsim 10^{-10}$ ph cm$^{-2}$ s$^{-1}$ integrated over a $5^{\circ}$ elongation angle and above 100 MeV: $\kappa$ Ori, $\zeta$ Pup, $\zeta$ Ori, Betelgeuse, $\delta$ Ori, Rigel, $\zeta$ Per, $\lambda$ Ori, and $\epsilon$ CMa. 

The IC halo emissions have been computed assuming a cosmic-ray electron spectrum similar to that observed at Earth and locally in the Galaxy \citep{orlando2018}. Although in stars like the Sun three components significantly contribute to the total $\gamma$-ray output, i.e., the stellar disk, the IC halo, and flares, in super-luminous stars the IC component is expected to outshine the other two (see \S \ref{sec:discussion} for details). After a decade of observations with {\it Fermi}-LAT, each star in our sample could in principle be detectable in $\gamma$-rays, at least at a marginal level ($\lesssim 3\sigma$). Indeed, there are several $\gamma$-ray sources listed in 4FGL with fluxes $< 10^{-10}$ ph cm$^{-2}$ s$^{-1}$ for energies $> 100$ MeV, including one extended source \citep{abdollahi2020_4FGL, ballet2020_4FGL-DR2}.

\section{IC models for single stars}
\label{sec:IC_models}

IC models for single stars follow the same formalism as the IC model of the Sun, which has been confirmed by $\gamma$-ray observations \citep{orlando2008gamma, abdo2011fermi}. The same formalism of IC emission has also been used for the interpretation of freshly accelerated cosmic rays in the Cygnus OB2 region \citep{ackermann2011Cygnus_Superbubble}, where a model of IC emission for 1700 O and B stars based on \cite{orlando2007gamma} was included in the analysis. 

Models for the 9 candidates described in Section \ref{sec:sample} are here updated with respect to the original estimations by accounting for more precise cosmic-ray electron measurements with the Alpha Magnetic Spectrometer \citep[AMS-02;][]{aguilar2014electron} on the International Space Station. IC models for each single source are calculated with the \texttt{StellarICS} code\footnote{\url{https://gitlab.mpcdf.mpg.de/aws/stellarics/}} \citep{orlando2020stellarics} following the formulation described in that paper and references therein. \texttt{StellarICS} computes the IC scattering in the heliosphere and photosphere of individual stars for a given cosmic-ray electron spectrum and modulation, and for a given photon spectral and spatial distribution. 

The IC emission for the single stars is computed assuming a cosmic-ray electron spectrum as observed at Earth after correcting for a modulation similar to the solar modulation. We adopt the same electron spectrum that fits AMS-02 cosmic-ray data \citep{aguilar2014electron} at 1~AU distance from the star with 600~MV demodulation potential to obtain the local interstellar spectrum. This means that we assume the local interstellar spectrum to be representative of the spectrum in the interstellar space, while assuming a fairly strong cosmic-ray modulation in the vicinity of the star. This also means that the IC emission could be greater than we estimate, due to electrons accelerated in stellar winds. Technical details on the stellar cosmic-ray electron spectrum and the IC modeling are given in \cite{orlando2020stellarics}. 

We produce two models to account for possible modulation of the electron spectrum due to stellar winds: 1) an extreme model that assumes that cosmic-ray electrons can not penetrate the stellar photosphere closer to 0.1$^{\circ}$ angular separation from the line-of-sight of the star, which for a star at 300 pc distance means 0.5~pc, and 2) a less extreme model that assumes that the cosmic-ray electrons can not penetrate the stellar photosphere closer to 0.025$^{\circ}$ angular separation from the line-of-sight of the star, which for a star at 300 pc distance means 0.13~pc. In the hypothesis that there is additional IC emission even closer to the star, a point-like component can account for it together with hadronic emission component. We tested that differences between the extreme and less extreme cases in the IC modeling do not significantly affect the results.

\begin{figure}
    \centering
    \includegraphics[width=\linewidth]{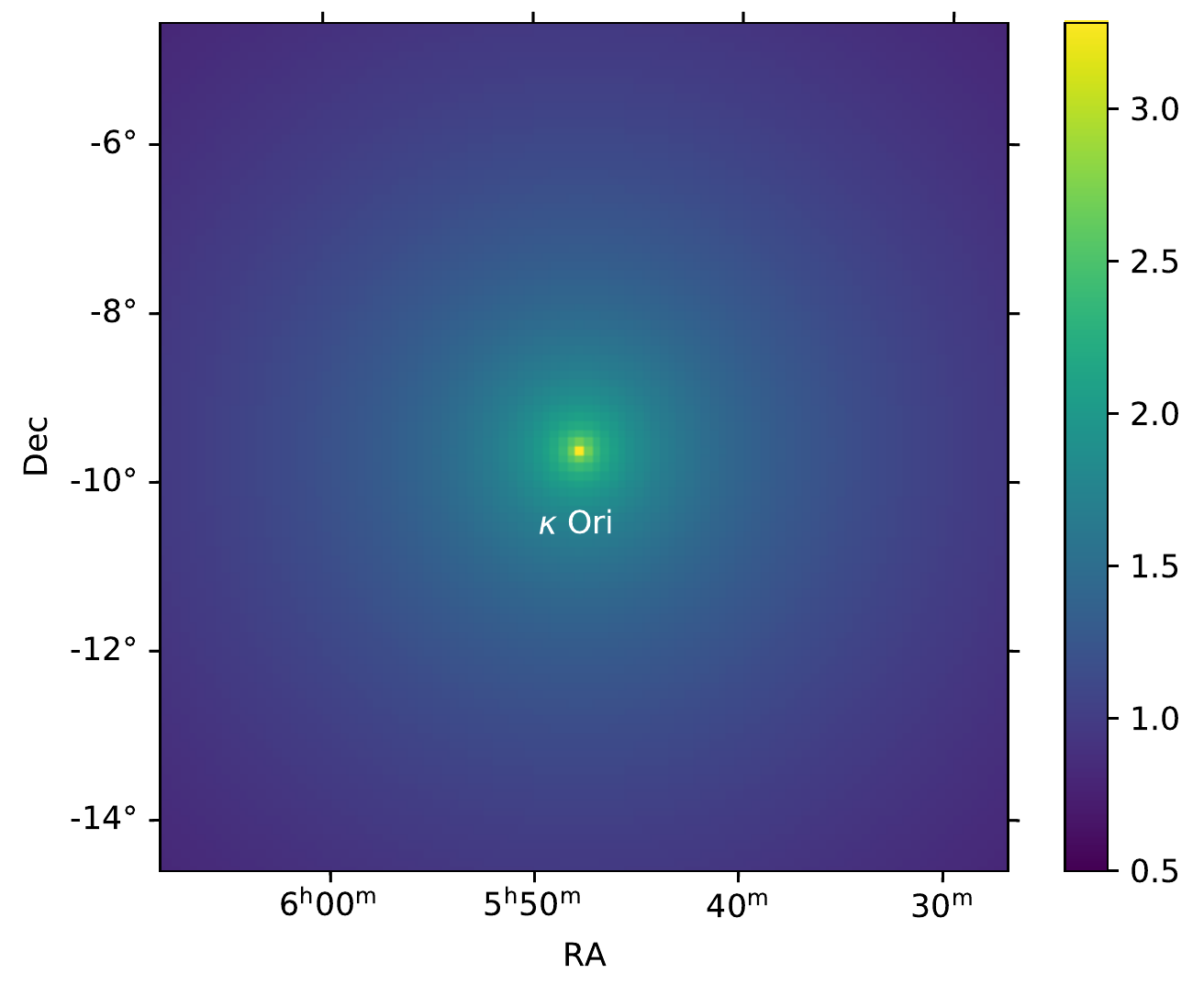}
    \caption{Extended IC halo model for $\kappa$ Ori. This extended map has a power-law spectrum and is normalized such that its integrated flux is equal to 1. The color bar is in log scale.}
    \label{fig:halo_model}
\end{figure}

In Figure \ref{fig:halo_model} we show the normalized IC extended emission map for $\kappa$~Ori, representing a region of 8$^{\circ}$ around the star, where we adopt a distance $d = 198$~pc, a surface temperature $T = 26500$~K, and a radius $r = 22.2$ R$_\odot$. The color map is in log scale and normalized according to \textit{Fermi}-LAT guidelines\footnote{\url{https://fermi.gsfc.nasa.gov/ssc/data/analysis/scitools/extended/extended.html}} such that its integrated flux is equal to 1. As mentioned in the previous paragraph, to account for a more stringent cosmic-ray modulation close to the star, we also test a secondary extended emission model where the central pixel of $0.1^{\circ} \times 0.1^{\circ}$ is very dim, however, we do not find any significant difference in the results. Additionally to the emission map of Figure \ref{fig:halo_model}, we also tested a mapcube model (i.e., an extended energy-dependent map) from \texttt{StellarICS} with spectral index fixed by the spectrum of AMS-02 electrons, but no significant improvement in the fitting was achieved.

\section{Observations and analysis}
\label{sec:observations}

For each star in our sample we select {\it Fermi}-LAT data within a $16^{\circ} \times 16^{\circ}$ region-of-interest (ROI) centered on the star positions given by \cite{van2007_Hipparcos_positions}. The period of observations spans nearly 12 years, ranging from August 4, 2008 to August 10, 2020, and the chosen energy range is 100 MeV -- 100 GeV (except for the stacking analysis, see \S \ref{subsec:observations_stacking}), divided into 8 logarithmically spaced bins per energy decade.

For each target in the sample we test i) a point-like model, for which we select \textit{Fermi}-LAT Pass 8 events belonging to the \texttt{SOURCE} class (\textit{evclass} $= 128$), which is the recommended class for analysing point-like and moderately extended sources; and ii) an extended stellar halo model (see Figure \ref{fig:halo_model}) based on the predicted IC emission described in \cite{orlando2007gamma}, for which we select {\it Fermi}-LAT Pass 8 events belonging to the \texttt{SOURCEVETO} class (\textit{evclass} $= 2048$), which is a class recommended for the analysis of diffuse $\gamma$-ray emission that requires low levels of cosmic-ray contamination. For both approaches, we use events converting in the front and back sections of the LAT tracker (\textit{evtype} $= 3$). We start the analysis adopting the curved spectra provided by the models computed with \texttt{StellarICS}, but our attempts of fitting the observed data failed due to low statistics. We instead decided to use a power-law spectrum for all targets, which has fewer independent parameters and is still suited for our case.

The $\gamma$-ray data are filtered for good time intervals with \texttt{DATA\_QUAL} $> 0$ and the recommended instrument configuration for science \texttt{LAT\_CONFIG} $== 1$, while the maximum zenith angle cut is set to $90^{\circ}$ to reduce contamination from the Earth limb. The Galactic and extragalactic background emissions are modeled with the latest interstellar emission model\footnote{\url{https://fermi.gsfc.nasa.gov/ssc/data/access/lat/BackgroundModels.html}} \textit{gll\_iem\_v07} and with the isotropic spectral templates \textit{iso\_P8R3\_SOURCE\_V2\_v1} or \textit{iso\_P8R3\_SOURCEVETO\_V2\_v1}, depending on the adopted data class.

The data analysis is performed with \texttt{Fermitools}\footnote{\url{https://fermi.gsfc.nasa.gov/ssc/data/analysis/software/}} v1.2.23 and \texttt{fermipy}\footnote{\url{https://fermipy.readthedocs.io/en/latest/index.html}} v0.17.4 \citep{wood2017fermipy}, by means of a binned likelihood analysis and using MINUIT as minimizer. For modelling each ROI we consider $\gamma$-ray sources listed in 4FGL as well as all sources found with the \texttt{fermipy} function \texttt{find\_sources()}. The normalization parameters of all sources lying within a radius of $5^{\circ}$ from the center of the ROIs are left free to vary, and sources listed in 4FGL and lying up to $5^{\circ}$ outside the ROIs have also been taken into account. To quantify the significance of our observations, we use a test statistic (TS) defined as $TS = 2(\mathcal{L}_1-\mathcal{L}_0)$, where the term inside parentheses is the difference between the maximum log-likelihoods with ($\mathcal{L}_1$) and without ($\mathcal{L}_0$) including our target in the model. The corresponding significance, in $\sigma$, is approximated by $\sqrt{TS}$ \citep{mattox1996likelihood}.

\subsection{Stacking analysis}
\label{subsec:observations_stacking}

Stacking $\gamma$-ray data is a powerful tool when we can characterize a population of sources with one or more average parameters. Although powerful, this method needs to be used with caution. If sources in the sample have very different characteristics from each other, the small contribution from each source will not sum up properly and the final result will be meaningless. In our case, we assume that the 9 stars in our sample can be characterized by an average photon index, $\Gamma$.

In an attempt to detect the collective emission of the 9 super-luminous stars, we perform a stacking analysis by summing up the log-likelihood values of each individual ROI, under the assumption that the stellar population can be characterized by the same photon index, $\Gamma$. The significance of the stacking, $TS_{stack}$, is independently computed over 41 photon index bins, $\Gamma_j$, ranging from $\Gamma_{j=0} = 1$ to $\Gamma_{j=40} = 5$ in intervals of $\Delta\Gamma = 0.1$, as given by the following equation
\begin{equation}
    TS_{stack,j} \left. \equiv 2\sum\limits_{i=1}^9 (\mathcal{L}_{1,i}-\mathcal{L}_{0,i}) \right|_{\Gamma_j},
    \label{eq:TS_stack}
\end{equation}
where $i$ runs over the 9 analysed ROIs, and $\mathcal{L}_1$ and $\mathcal{L}_0$ have been defined in the previous section. The final result is a distribution of $TS_{stack}$ in terms of $\Gamma$ (see \S \ref{sec:resultados_stacking}).

To generate the log-likelihood profiles, we first perform a standard fit on each ROI, similar to what is described in the previous section, where we consider the normalization of all sources lying up to a radius of $5^{\circ}$ from the ROI center free to vary. The exceptions are the isotropic and Galactic emission components, which we leave completely free. After that, we fix the parameters of all point-like sources at their optimized values, add our target to the model with a fixed photon index $\Gamma_j$ and free normalization, and then run the fit again over the 41 values of $\Gamma_j$. To compare our results with background fluctuations, we repeat the exact same stacking procedure for 12 mock sources added to the neighborhood of each star (i.e., 108 mock sources in total) and take the average distribution of $TS_{stack}$ in terms of $\Gamma$ (see \S \ref{sec:resultados_stacking}). The mock sources are added to the model in order to compute random background fluctuations that could mimic the presence of real faint sources with the same properties of our targets.


One major difference of the stacking analysis shown here, when compared with the analysis presented in the previous section, is that we restrict the energy range to the window of 500 MeV -- 100 GeV. This cut is made because we noticed that small imprecision due to inaccurate modelling of the interstellar $\gamma$-ray emission could leave behind a substantially high number of unmodeled soft photons, which can in turn be absorbed into the model of our target, especially when we are dealing with photon indexes $\Gamma \gtrsim 3.5$, resulting in unrealistic TS values. This problem is mitigated when we cut off the low-energy photons (i.e., $< 500$ MeV). For more details on stacking {\it Fermi}-LAT data, we refer the reader to \cite{paliya2019fermi}.


\section{Results}
\label{sec:resultados}

The energy flux upper limits for all 9 stars in our sample are shown in Table \ref{tab:results} along with their predicted extended halo IC fluxes integrated over elongation angles of $1^{\circ}$ and $8^{\circ}$. To compute these upper limits, we assume that the IC halos can be well described by a power-law spectrum with photon index $\Gamma = 2$, which is a reasonable choice if we expect this emission to be similar to that observed in the Sun \citep{orlando2008gamma,abdo2011fermi}. For simplicity, we also assume that the integrated IC emission up to $1^{\circ}$ elongation angle can be well represented by a point-like $\gamma$-ray source in our ROI models. In Table \ref{tab:results} we also show the Galactic coordinates of each star as well as the Galactic $\gamma$-ray fluxes integrated over elongation angles of $1^{\circ}$ and $8^{\circ}$.


The predicted IC emission for all of the stars in our sample, with the exception of $\zeta$ Puppis, is consistent with the 95\% confidence level flux upper limits in at least one of the two (point-like and extended) adopted models. To test the detectability of $\zeta$ Puppis, we use the \texttt{fermipy} function \texttt{simulate\_source()} to simulate $\gamma$-ray data representing the stellar halo of $\zeta$ Puppis. We assume the halo to have a power-law spectrum with $\Gamma = 2$ and an integrated flux starting at $3.1 \times 10^{-10}$ cm$^{-2}$ s$^{-1}$, which is the predicted IC flux for $\zeta$ Puppis $> 100$ MeV and integrated over $1^{\circ}$. The resulting TS for this simulated source was $\sim 1.8$. We then gradually increase the simulated flux in steps of $3.1 \times 10^{-10}$ cm$^{-2}$ s$^{-1}$ until we get a $3\sigma$ (TS $= 9$) detection with \textit{Fermi}-LAT. The simulation shows that the IC flux of $\zeta$ Puppis should be at least $\sim 4$ times higher than predicted in order to be detected by \textit{Fermi}-LAT at the $3\sigma$ confidence level.




\begin{table*}
    \centering
    \begin{tabular}{l|c|c|c|c|c|c|c|c}
        Name & Pred. flux $1^{\circ}$ & UL PM & Pred. flux $8^{\circ}$ & UL EM & l & b & Gal. flux $1^{\circ}$ & Gal. flux $8^{\circ}$ \\
           & ($10^{-10}$ cm$^{-2}$ s$^{-1}$) & ($10^{-10}$ cm$^{-2}$ s$^{-1}$) & ($10^{-10}$ cm$^{-2}$ s$^{-1}$) & ($10^{-10}$ cm$^{-2}$ s$^{-1}$) & ($^{\circ}$) & ($^{\circ}$) & ($10^{-6}$ cm$^{-2}$ s$^{-1}$) & ($10^{-4}$ cm$^{-2}$ s$^{-1}$)\\
        \hline
        $\kappa$ Ori & 1.6 & 5.4 & 13 & 34 & 214.514 & -18.496 & $5.9\pm0.2$ & $3.75\pm0.01$ \\
        Betelgeuse & 2.3 & 4.1 & 17 & 23  & 199.787 & -8.959 & $5.8\pm0.2$  & $3.74\pm0.01$ \\
        $\delta$ Ori & 0.6 & 2.7 & 3.7 & 14  & 203.856 & -17.740 & $5.3\pm0.1$ & $3.39\pm0.01$ \\
        $\epsilon$ CMa & 0.7 & 2.4 & 5.5 & 20  & 239.831 & -11.330 & $6.2\pm0.1$ & $3.93\pm0.01$ \\
        $\lambda$ Ori & 0.4 & 4.2 & 2.4 & 27  & 195.052 & -11.995 & $6.1\pm0.2$ & $3.87\pm0.01$ \\
        Rigel & 1.6 & 3.3 & 13 & 21  & 209.241 & -25.245 & $5.7\pm0.2$ & $3.65\pm0.01$ \\
        $\zeta$ Ori & 0.9 & 0.9 & 6.4 & 11  & 206.452 & -16.585 & $5.9\pm0.1$ & $3.76\pm0.01$ \\
        $\zeta$ Per & 1.1 & 4.4 & 9.2 & 25  & 162.289 & -16.690 & $5.8\pm0.2$ & $3.67\pm0.02$ \\
        $\zeta$ Pup & 3.1 & 3.0 & 26 & 9.5  & 255.976 & -4.706 & $6.26\pm0.05$ & $3.986\pm0.004$ 
    \end{tabular}
    \caption{Predicted IC $\gamma$-ray fluxes integrated over $1^{\circ}$ and $8^{\circ}$ elongation angles, and 95\% confidence level upper limits (UL) for energies $> 100$ MeV and photon index $\Gamma = 2$. PM stands for ``point-like model'', and EM stands for ``extended model''. $\zeta$ Puppis is the only star for which the upper limits are not consistent with the theoretical predictions. The TSs for each individual target are nearly zero, independently of the adopted model. In the last columns we show the Galactic coordinates of each star and the measured Galactic fluxes integrated over $1^{\circ}$ and $8^{\circ}$ elongation angles.}
    \label{tab:results}
\end{table*}

\begin{figure*}
    \centering
    \includegraphics[scale=0.50]{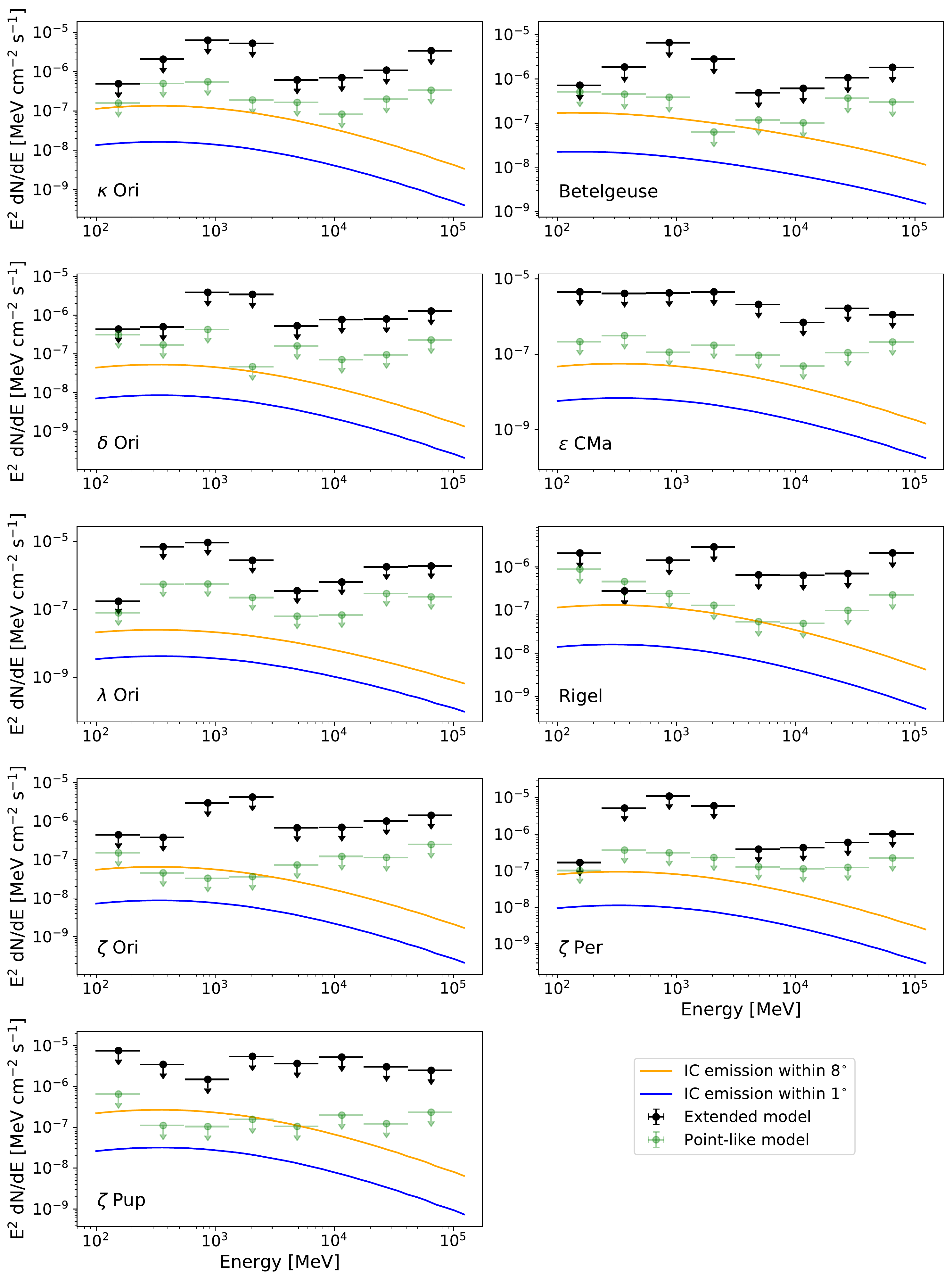}
    \caption{SEDs for all stars in our sample. The black and green dots represent the 95\% confidence level upper limits when we model the targets as extended $\gamma$-ray halos and point-like sources, respectively. The solid lines represent the predicted IC halo SEDs integrated over $1^{\circ}$ (blue) and $8^{\circ}$ (orange). We see that for $\zeta$ Pup, $\zeta$ Ori, Rigel and $\zeta$ Per, some upper limits are barely consistent with the models, suggesting that the detection of the sample as a population could be feasible via stacking analysis.}
    \label{fig:SEDs}
\end{figure*}

The stellar spectral energy distributions (SEDs) are shown in Figure \ref{fig:SEDs}, where we plot the theoretical IC spectra integrated over elongation angles of $1^{\circ}$ (blue) and $8^{\circ}$ (orange) as solid lines, overlaid with the upper limits obtained when modeling the targets as extended sources (black dots) and point-like sources (green dots). We see that in general the upper limits are in agreement with the predictions. However, for a few sources like $\zeta$ Orionis and $\zeta$ Puppis, some upper limits are barely consistent with the models, raising the possibility of a global detection via stacking analysis.


\subsection{Results from stacking}
\label{sec:resultados_stacking}

Even if we could not detect the individual stars in our previous analysis, the sum of their predicted IC fluxes $> 500$ MeV (see \S \ref{subsec:observations_stacking} for energy cuts) reach values of $1.86$ and $14.7$ $\times 10^{-10}$ ph cm$^{-2}$ s$^{-1}$ when integrated over elongation angles of $1^{\circ}$ and $8^{\circ}$, respectively. These values are very appealing for a stacking analysis, as other authors have already measured \textit{Fermi}-LAT stacked fluxes as low as $6 \times 10^{-12}$ ph cm$^{-2}$ s$^{-1}$ \citep{paliya2019fermi}, although they were working with blazars on an energy range $> 10$ GeV.

\begin{figure*}
    \centering
    \includegraphics[width=\linewidth]{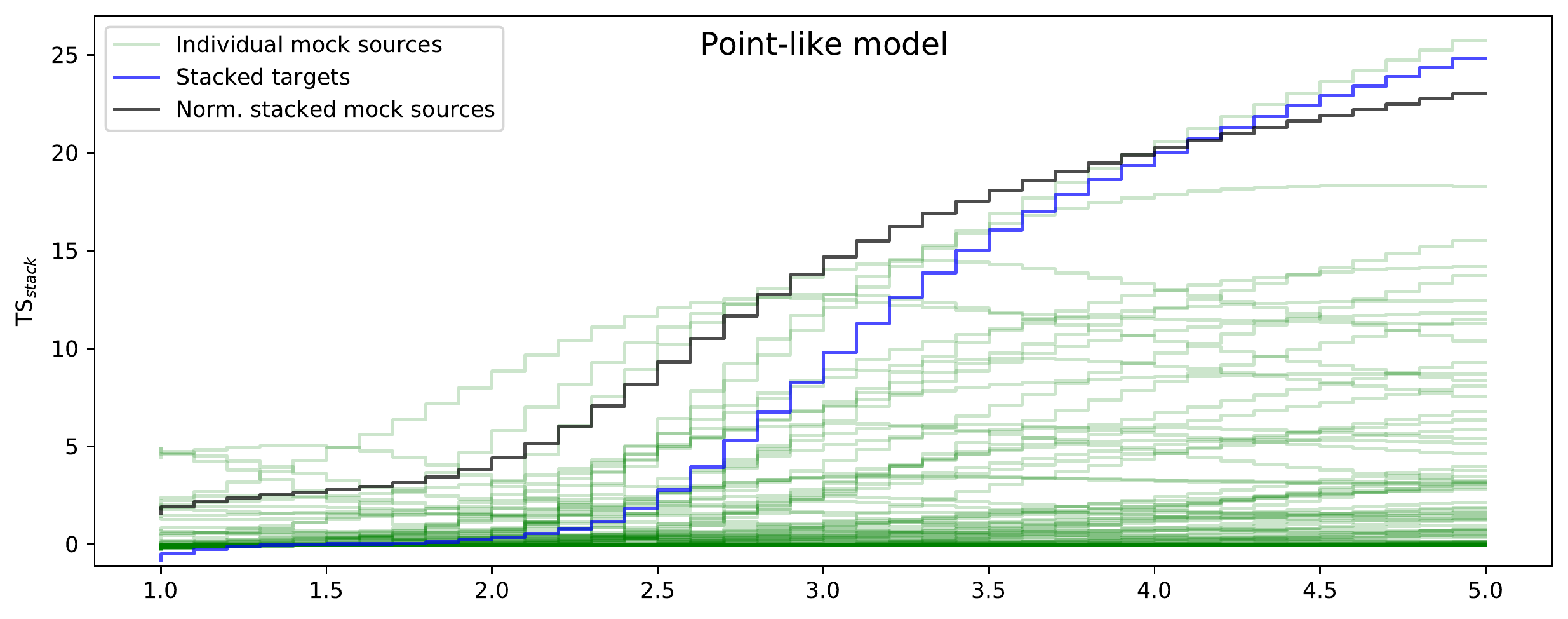}
    \includegraphics[width=\linewidth]{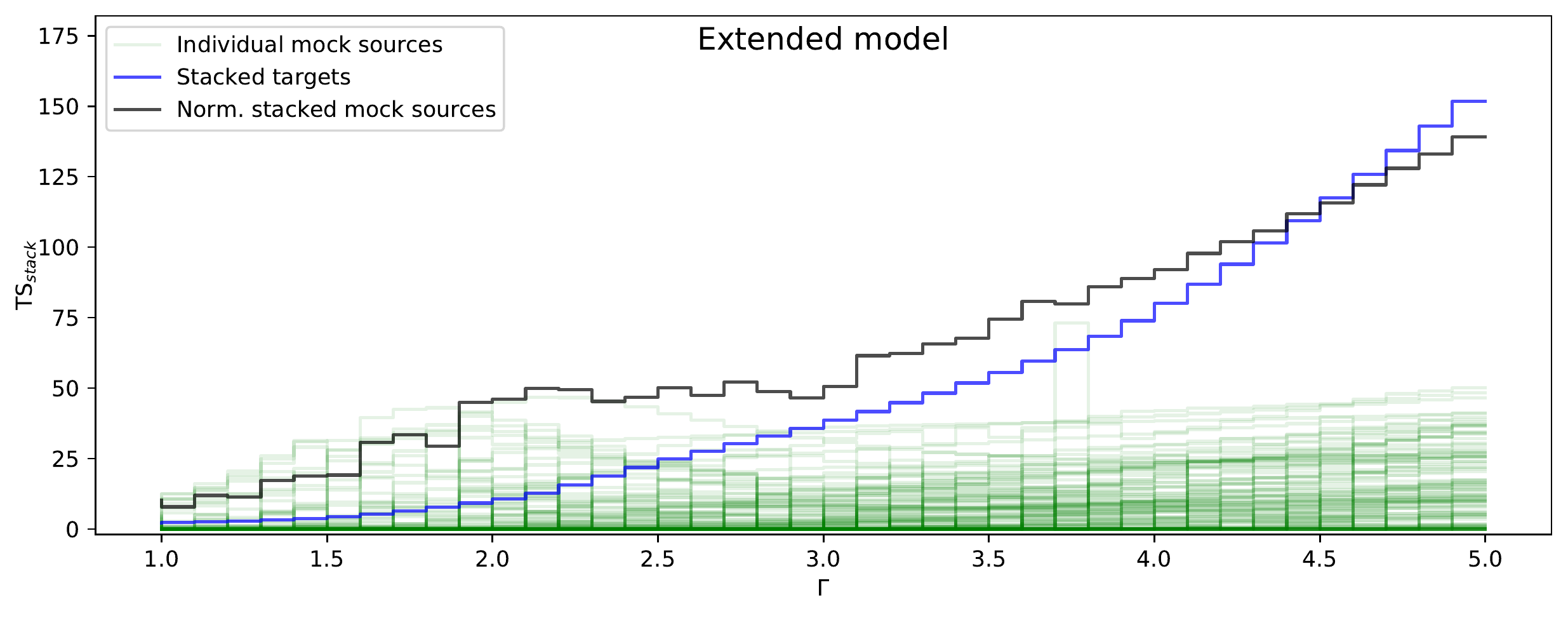}
    \caption{Distribution of TS$_{stack}$ in terms of the photon index $\Gamma$. The blue line represents the stacked TS for our targets, while the black line represents the normalized stacked background, built with the stacking of 108 mock sources (green lines). Our results are consistent with random background fluctuations. The difference between both panels is that in the top we model the targets as a point sources, while in the bottom we use the extended map shown in \S \ref{sec:observations}.}
    \label{fig:stacking_pointsources}
\end{figure*}

Our results for the stacking analysis are shown in Figure \ref{fig:stacking_pointsources} as a distribution of TS$_{stack}$ in terms of $\Gamma$. If there was a significant signal from the stacked stars, the blue line in the figure should be peaked around a specific value of $\Gamma$ and present values substantially higher than the stacked background (black line). In order to compute the background, we stack 108 mock sources (green lines) scattered at random positions in the vicinity (up to $6^{\circ}$) of our targets. Both panels in the figure represent exactly the same thing, except that for the top panel we model the targets as point-sources and in the lower panel we model them with the extended map described in \S \ref{sec:IC_models}. As is clear from the figure, our results are consistent with background noise, implying that the quiescent $\gamma$-ray emission from nearby super-luminous stars still cannot be detected by \textit{Fermi}-LAT.

\begin{figure}
    \centering
    \includegraphics[width=\linewidth]{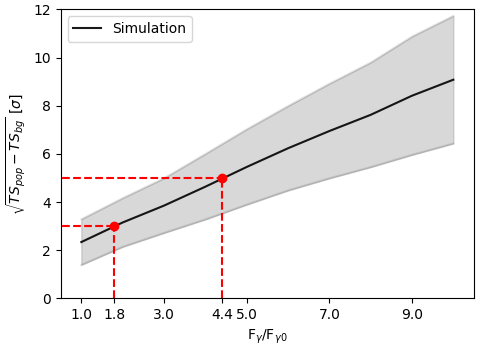}
    \caption{Stacked significance (black line) of a sample of simulated super-luminous stars in terms of the total predicted $\gamma$-ray flux of our original sample $F_{\gamma,0} = 1.86 \times 10^{-10}$ ph cm$^{-2}$ s$^{-1}$ ($> 500$ MeV and integrated over $1^{\circ}$ elongation angle). In the vertical axis, $TS_{pop}$ and $TS_{bg}$ stand for the TS of the simulated population and average background, respectively. The red dashed lines indicate that the population of stars should be at least 1.8 or 4.4 times brighter in order to be detected by \textit{Fermi}-LAT at the $3\sigma$ or $5\sigma$ confidence level, respectively. The gray band represents the standard deviation of the stacking, assuming that the stars have roughly the same $\gamma$-ray brightness.}
    \label{fig:simulation}
\end{figure}

To constrain the total $\gamma$-ray flux of our sample, we simulate 12 $\gamma$-ray stellar halos in the neighborhood of each target (a total of 108 simulated sources). We start by simulating halos with power-law index $\Gamma = 2$ and exactly the same IC luminosity predicted for our targets. We then gradually increase their luminosity up to 10 times its original value. As the summed $\gamma$-ray flux for the 9 stars is $F_{\gamma,0} = 1.86 \times 10^{-10}$ ph cm$^{-2}$ s$^{-1}$ ($> 500$ MeV and integrated up to $1^{\circ}$ elongation angle), the current significance of our sample should lie between 1.3 and $3.1\sigma$, as shown in Figure \ref{fig:simulation}. If our IC model correctly predicts the stellar fluxes, the population should be at least 4.4 times brighter in order to be detected with \textit{Fermi}-LAT at the $\sim 5\sigma$ confidence level. From the figure, we see that our sample must emit less than $F_{\gamma} = 1.8 \times F_{\gamma,0}$, otherwise it should be detected at the $3\sigma$ confidence level. At this significance level, each star should, on average, emit $< 3.3 \times 10^{-11}$ ph cm$^{-2}$ s$^{-1}$ ($> 500$ MeV and integrated over $1^{\circ}$ elongation angle), which is one order of magnitude more restrictive than the typical upper limit, $F_{\gamma,ul} \lesssim 3 \times 10^{-10}$ cm$^{-2}$ s$^{-1}$, that we get for the individual sources integrated over the same ranges of energy and elongation angle.

\section{Discussion}
\label{sec:discussion}

None of the stars in our sample have been detected in $\gamma$-rays, either individually or the population as a whole. Even if our models take only the IC halo component into account, this component should be dominant in super-luminous stars. However, the possibility of enhanced $\gamma$-ray flux due to massive stellar flares is also appealing \citep{ohm2018expected}. For the Sun, some massive flares emit more $\gamma$-rays within one hour than the whole solar disk in one year \citep{abdo2011fermi,Ackermann2014_solar_flares}. These flares usually have a soft photon index, typically $\Gamma > 2.5$ for power-law models \citep{ajello2021_fermi_solar_flare_cat}. As main sequence and giant stars are also known to host stellar flares \citep{pettersen1989review}, this could significantly boost their total flux below $\sim 500$ MeV (i.e., out of reach for our stacking analysis).

The $\gamma$-ray flux from the solar disk component ($\sim 4.6 \times 10^{-7}$ ph cm$^{-2}$ s$^{-1}$) is comparable to the flux of the solar IC component ($\sim 6.8 \times 10^{-7}$ ph cm$^{-2}$ s$^{-1}$) integrated over an elongation angle of $20^{\circ}$ \citep{abdo2011fermi}. Even if the disk component emission significantly contributes to the total $\gamma$-ray flux of the quiet Sun, a contribution from stellar disks of super-luminous stars should likely be negligible. Although there is no satisfactory theory about the $\gamma$-ray emission from cosmic-ray interactions with stellar atmospheres (as mentioned in \S \ref{sec:intro}), we can at least naively estimate how strong the disk components of super-luminous stars are by assuming that their $\gamma$-ray intensities may be of the same order of magnitude as the solar disk emission. In this sense, Betelgeuse should have the brightest disk component in our sample. With a radius of $764 R_{\odot}$ and distance of 168 pc \citep{Meridith2020_betelgeuse}, its disk component is (naively) $\sim 5.8\times 10^5$ brighter than the solar disk and $3.45 \times 10^7$ more distant, implying in a flux $\sim 10^{-10}$ that of the solar disk.

As discussed in the previous section, the stars in our sample could only be detected at the $5\sigma$ level by \textit{Fermi}-LAT if they were 4.4 times brighter, assuming our model for IC emission is correct. Given that in our models we consider local cosmic-ray electron spectral flux similar to that observed in the Solar System, and that the $\gamma$-ray emission from stellar halos is predicted to be directly proportional to the local density of cosmic-ray electrons \citep{orlando2007gamma}, our stacking analysis constrains the density of cosmic-ray electrons in the vicinity of our targets to be less than $\sim 4$ (or $\sim 2$) times the density observed in the Solar System, otherwise we should detect the stacked stars at the $5\sigma$ (or $3\sigma$) confidence level. Important sources of uncertainty in the $\gamma$-ray flux estimates come from the uncertainty in the estimated distances and luminosities of the stars, with a range of 5 to 20\% errors reported in the recent literature when available. In first approximation, these uncertainties apply linearly to the derived $\gamma$-ray fluxes of the stars.

\section{Conclusions}
\label{sec:conclusions}

In this work we used 12 years of \textit{Fermi}-LAT observations to perform the first systematic study of super-luminous stars in $\gamma$-rays, where we tested if \textit{Fermi}-LAT could detect the IC emission from extended stellar photon halos. Our main conclusions are summarized as follows:

\begin{itemize}
    \item We did not find significant detection of the IC emission from stellar halos around super-luminous stars.
    \item We found that the stars in our sample must emit on average $< 3.3 \times 10^{-11}$ ph cm$^{-2}$ s$^{-1}$ for energies $> 500$ MeV and integrating the IC halo over $1^{\circ}$ elongation angle, otherwise we should detect them as a population at the $3\sigma$ confidence level. This result constrains the density of cosmic-ray electrons in the surroundings of our targets to be less than twice the value we measure in the Solar System.
\end{itemize}

There is the possibility of detecting these stars with the upcoming All-sky Medium Energy Gamma-ray Observatory \citep[AMEGO;][]{McEnery2019_AMEGO}, which will be $2 \sim 10$ times more sensitive than \textit{Fermi}-LAT at energies $\lesssim 200$ MeV. At this energy range, AMEGO will have a 5 times better angular resolution than \textit{Fermi}-LAT, thus being able to better distinguish between diffuse and stellar emission.

\section*{Acknowledgements}

We thank Melissa Pesce-Rollins and David J. Thompson for the constructive comments allowing us to improve the manuscript. E.O. acknowledges the ASI-INAF agreement n. 2017-14-H.0 and the NASA Grant No. 80NSSC20K1558.

The \textit{Fermi}-LAT Collaboration acknowledges generous ongoing support
from a number of agencies and institutes that have supported both the
development and the operation of the LAT as well as scientific data analysis.
These include the National Aeronautics and Space Administration and the
Department of Energy in the United States, the Commissariat \`a l'Energie Atomique
and the Centre National de la Recherche Scientifique / Institut National de Physique
Nucl\'eaire et de Physique des Particules in France, the Agenzia Spaziale Italiana
and the Istituto Nazionale di Fisica Nucleare in Italy, the Ministry of Education,
Culture, Sports, Science and Technology (MEXT), High Energy Accelerator Research
Organization (KEK) and Japan Aerospace Exploration Agency (JAXA) in Japan, and
the K.~A.~Wallenberg Foundation, the Swedish Research Council and the
Swedish National Space Board in Sweden.

MDM research is supported by Fellini - Fellowship for Innovation at INFN, funded by the European Union’s Horizon 2020 research programme under the Marie Skłodowska-Curie Cofund Action, grant agreement no.~754496.
 
Additional support for science analysis during the operations phase is gratefully
acknowledged from the Istituto Nazionale di Astrofisica in Italy and the Centre
National d'\'Etudes Spatiales in France. This work performed in part under DOE
Contract DE-AC02-76SF00515.

\section*{Data Availability}

All data used in this work can be found online in the \textit{Fermi}-LAT data server\footnote{\url{https://fermi.gsfc.nasa.gov/cgi-bin/ssc/LAT/LATDataQuery.cgi}}. The data analysis was performed with \texttt{Fermitools} and \texttt{fermipy}, both software free to download and use (see \S \ref{sec:observations}).



\bibliographystyle{mnras}
\bibliography{main} 

\begin{thebibliography}{}
\makeatletter
\relax
\def\mn@urlcharsother{\let\do\@makeother \do\$\do\&\do\#\do\^\do\_\do\%\do\~}
\def\mn@doi{\begingroup\mn@urlcharsother \@ifnextchar [ {\mn@doi@}
  {\mn@doi@[]}}
\def\mn@doi@[#1]#2{\def\@tempa{#1}\ifx\@tempa\@empty \href
  {http://dx.doi.org/#2} {doi:#2}\else \href {http://dx.doi.org/#2} {#1}\fi
  \endgroup}
\def\mn@eprint#1#2{\mn@eprint@#1:#2::\@nil}
\def\mn@eprint@arXiv#1{\href {http://arxiv.org/abs/#1} {{\tt arXiv:#1}}}
\def\mn@eprint@dblp#1{\href {http://dblp.uni-trier.de/rec/bibtex/#1.xml}
  {dblp:#1}}
\def\mn@eprint@#1:#2:#3:#4\@nil{\def\@tempa {#1}\def\@tempb {#2}\def\@tempc
  {#3}\ifx \@tempc \@empty \let \@tempc \@tempb \let \@tempb \@tempa \fi \ifx
  \@tempb \@empty \def\@tempb {arXiv}\fi \@ifundefined
  {mn@eprint@\@tempb}{\@tempb:\@tempc}{\expandafter \expandafter \csname
  mn@eprint@\@tempb\endcsname \expandafter{\@tempc}}}

\bibitem[\protect\citeauthoryear{Abdo et~al.,}{Abdo
  et~al.}{2011}]{abdo2011fermi}
Abdo A.,  et~al., 2011, The Astrophysical Journal, 734, 116

\bibitem[\protect\citeauthoryear{Abdo et~al.,}{Abdo
  et~al.}{2012}]{abdo2012_Moon_fermi-LAT}
Abdo A.~A.,  et~al., 2012, The Astrophysical Journal, 758, 140

\bibitem[\protect\citeauthoryear{Abdollahi et~al.,}{Abdollahi
  et~al.}{2020}]{abdollahi2020_4FGL}
Abdollahi S.,  et~al., 2020, The Astrophysical Journal Supplement Series, 247,
  33

\bibitem[\protect\citeauthoryear{{Ackermann} et~al.,}{{Ackermann}
  et~al.}{2011}]{ackermann2011Cygnus_Superbubble}
{Ackermann} M.,  et~al., 2011, \mn@doi [Science] {10.1126/science.1210311},
  \href {https://ui.adsabs.harvard.edu/abs/2011Sci...334.1103A} {334, 1103}

\bibitem[\protect\citeauthoryear{{Ackermann} et~al.,}{{Ackermann}
  et~al.}{2014}]{Ackermann2014_solar_flares}
{Ackermann} M.,  et~al., 2014, \mn@doi [\apj] {10.1088/0004-637X/787/1/15},
  \href {https://ui.adsabs.harvard.edu/abs/2014ApJ...787...15A} {787, 15}

\bibitem[\protect\citeauthoryear{Aguilar et~al.,}{Aguilar
  et~al.}{2014}]{aguilar2014electron}
Aguilar M.,  et~al., 2014, Physical review letters, 113, 121102

\bibitem[\protect\citeauthoryear{Ajello et~al.,}{Ajello
  et~al.}{2021}]{ajello2021_fermi_solar_flare_cat}
Ajello M.,  et~al., 2021, The Astrophysical Journal Supplement Series, 252, 13

\bibitem[\protect\citeauthoryear{{Ballet}, {Burnett}, {Digel}  \&
  {Lott}}{{Ballet} et~al.}{2020}]{ballet2020_4FGL-DR2}
{Ballet} J.,  {Burnett} T.~H.,  {Digel} S.~W.,   {Lott} B.,  2020, arXiv
  e-prints, \href {https://ui.adsabs.harvard.edu/abs/2020arXiv200511208B} {p.
  arXiv:2005.11208}

\bibitem[\protect\citeauthoryear{{ESA}}{{ESA}}{1997}]{esa1997_Hipp_Tycho_cats}
{ESA} .,  1997, VizieR Online Data Catalog, \href
  {https://ui.adsabs.harvard.edu/abs/1997yCat.1239....0E} {p. I/239}

\bibitem[\protect\citeauthoryear{Guti{\'e}rrez \& Masip}{Guti{\'e}rrez \&
  Masip}{2020}]{gutierrez2020sun}
Guti{\'e}rrez M.,  Masip M.,  2020, Astroparticle Physics, 119, 102440

\bibitem[\protect\citeauthoryear{{Joyce}, {Leung}, {Moln{\'a}r}, {Ireland},
  {Kobayashi}  \& {Nomoto}}{{Joyce} et~al.}{2020}]{Meridith2020_betelgeuse}
{Joyce} M.,  {Leung} S.-C.,  {Moln{\'a}r} L.,  {Ireland} M.,  {Kobayashi} C.,
  {Nomoto} K.,  2020, \mn@doi [\apj] {10.3847/1538-4357/abb8db}, \href
  {https://ui.adsabs.harvard.edu/abs/2020ApJ...902...63J} {902, 63}

\bibitem[\protect\citeauthoryear{Kanbach et~al.,}{Kanbach
  et~al.}{1989}]{kanbach1989project}
Kanbach G.,  et~al., 1989, Space Science Reviews, 49, 69

\bibitem[\protect\citeauthoryear{{Li}, {Ng}, {Chen}, {Nan}  \& {He}}{{Li}
  et~al.}{2020}]{li2020simulating}
{Li} Z.,  {Ng} K. C.~Y.,  {Chen} S.,  {Nan} Y.,   {He} H.,  2020, arXiv
  e-prints, \href {https://ui.adsabs.harvard.edu/abs/2020arXiv200903888L} {p.
  arXiv:2009.03888}

\bibitem[\protect\citeauthoryear{Linden, Zhou, Beacom, Peter, Ng  \&
  Tang}{Linden et~al.}{2018}]{linden2018evidence}
Linden T.,  Zhou B.,  Beacom J.~F.,  Peter A.~H.,  Ng K.~C.,   Tang Q.-W.,
  2018, Physical review letters, 121, 131103

\bibitem[\protect\citeauthoryear{Mattox et~al.,}{Mattox
  et~al.}{1996}]{mattox1996likelihood}
Mattox J.~R.,  et~al., 1996, The Astrophysical Journal, 461, 396

\bibitem[\protect\citeauthoryear{Mazziotta, {Luque}, {Di Venere}, {Fass{\`o}},
  {Ferrari}, {Loparco}, {Sala}  \& {Serini}}{Mazziotta
  et~al.}{2020a}]{mazziotta2020_FLUKA_code}
Mazziotta M.,  {Luque} P. D. L.~T.,  {Di Venere} L.,  {Fass{\`o}} A.,
  {Ferrari} A.,  {Loparco} F.,  {Sala} P.~R.,   {Serini} D.,  2020a, \mn@doi
  [\prd] {10.1103/PhysRevD.101.083011}, \href
  {https://ui.adsabs.harvard.edu/abs/2020PhRvD.101h3011M} {101, 083011}

\bibitem[\protect\citeauthoryear{Mazziotta, Loparco, Serini, Cuoco, Luque,
  Gargano  \& Gustafsson}{Mazziotta et~al.}{2020b}]{mazziotta2020search}
Mazziotta M.,  Loparco F.,  Serini D.,  Cuoco A.,  Luque P. D. L.~T.,  Gargano
  F.,   Gustafsson M.,  2020b, Physical Review D, 102, 022003

\bibitem[\protect\citeauthoryear{{McEnery} et~al.,}{{McEnery}
  et~al.}{2019}]{McEnery2019_AMEGO}
{McEnery} J.,  et~al., 2019, in Bulletin of the American Astronomical Society.
  p.~245 (\mn@eprint {arXiv} {1907.07558})

\bibitem[\protect\citeauthoryear{Moskalenko, Porter  \& Digel}{Moskalenko
  et~al.}{2006}]{moskalenko2006inverse}
Moskalenko I.~V.,  Porter T.~A.,   Digel S.~W.,  2006, The Astrophysical
  Journal Letters, 652, L65

\bibitem[\protect\citeauthoryear{Niblaeus, Beniwal  \& Edsj{\"o}}{Niblaeus
  et~al.}{2019}]{niblaeus2019neutrinos}
Niblaeus C.,  Beniwal A.,   Edsj{\"o} J.,  2019, Journal of Cosmology and
  Astroparticle Physics, 2019, 011

\bibitem[\protect\citeauthoryear{Ohm \& Hoischen}{Ohm \&
  Hoischen}{2018}]{ohm2018expected}
Ohm S.,  Hoischen C.,  2018, Monthly Notices of the Royal Astronomical Society,
  474, 1335

\bibitem[\protect\citeauthoryear{{Orlando}}{{Orlando}}{2018}]{orlando2018}
{Orlando} E.,  2018, \mn@doi [\mnras] {10.1093/mnras/stx3280}, \href
  {https://ui.adsabs.harvard.edu/abs/2018MNRAS.475.2724O} {475, 2724}

\bibitem[\protect\citeauthoryear{Orlando \& Strong}{Orlando \&
  Strong}{2007}]{orlando2007gamma}
Orlando E.,  Strong A.,  2007, \mn@doi [Astrophysics and Space Science]
  {10.1007/s10509-007-9457-0}, \href
  {https://ui.adsabs.harvard.edu/abs/2007Ap&SS.309..359O} {309, 359}

\bibitem[\protect\citeauthoryear{Orlando \& Strong}{Orlando \&
  Strong}{2008}]{orlando2008gamma}
Orlando E.,  Strong A.~W.,  2008, Astronomy \& Astrophysics, 480, 847

\bibitem[\protect\citeauthoryear{Orlando \& Strong}{Orlando \&
  Strong}{2020}]{orlando2020stellarics}
Orlando E.,  Strong A.,  2020, arXiv preprint arXiv:2012.13126

\bibitem[\protect\citeauthoryear{Paliya, Dominguez, Ajello, Franckowiak  \&
  Hartmann}{Paliya et~al.}{2019}]{paliya2019fermi}
Paliya V.~S.,  Dominguez A.,  Ajello M.,  Franckowiak A.,   Hartmann D.,  2019,
  The Astrophysical Journal Letters, 882, L3

\bibitem[\protect\citeauthoryear{Pettersen}{Pettersen}{1989}]{pettersen1989review}
Pettersen B.,  1989, in International Astronomical Union Colloquium. pp
  299--312

\bibitem[\protect\citeauthoryear{Riley, Strigari, Porter, Blandford, Murgia,
  Kerr  \& J{\'o}hannesson}{Riley et~al.}{2019}]{riley2019possible}
Riley A.~H.,  Strigari L.~E.,  Porter T.~A.,  Blandford R.~D.,  Murgia S.,
  Kerr M.,   J{\'o}hannesson G.,  2019, The Astrophysical Journal, 878, 8

\bibitem[\protect\citeauthoryear{Seckel, Stanev  \& Gaisser}{Seckel
  et~al.}{1991}]{seckel1991signatures}
Seckel D.,  Stanev T.,   Gaisser T.,  1991, The Astrophysical Journal, 382, 652

\bibitem[\protect\citeauthoryear{Song \& Paglione}{Song \&
  Paglione}{2020}]{song2020stacking}
Song Y.,  Paglione T.~A.,  2020, The Astrophysical Journal, 900, 185

\bibitem[\protect\citeauthoryear{Tang, Ng, Linden, Zhou, Beacom  \& Peter}{Tang
  et~al.}{2018}]{tang2018unexpected}
Tang Q.-W.,  Ng K.~C.,  Linden T.,  Zhou B.,  Beacom J.~F.,   Peter A.~H.,
  2018, Physical Review D, 98, 063019

\bibitem[\protect\citeauthoryear{Van~Leeuwen}{Van~Leeuwen}{2007}]{van2007_Hipparcos_positions}
Van~Leeuwen F.,  2007, Astronomy \& Astrophysics, 474, 653

\bibitem[\protect\citeauthoryear{Wood, Caputo, Charles, Di~Mauro, Magill  \&
  Perkins}{Wood et~al.}{2017}]{wood2017fermipy}
Wood M.,  Caputo R.,  Charles E.,  Di~Mauro M.,  Magill J.,   Perkins J.,
  2017, arXiv preprint arXiv:1707.09551

\makeatother
\end{thebibliography}









\bsp	
\label{lastpage}
\end{document}